\documentstyle[psfig,astrobib]{mn-ab}
\newif\ifAMStwofonts
\AMStwofontstrue

\def\kms{km\,s$^{-1}$}
\def\feii{[Fe\,II]}
\def\h2{H$_2$}

\def\mic{$\,\mu$m}

\title{An infrared proper motion study of the Orion bullets}
\author[J.-K. Lee \& M.\ G. Burton]
       {J.-K.~Lee\thanks{email: jklee@phys.unsw.edu.au} and M.\ G.~Burton\\
        School of Physics, University of New South Wales, Sydney, NSW 2052,
        Australia}
\date{Submitted: 3 August 1999}

\pagerange{\pageref{firstpage}--\pageref{lastpage}}
\pubyear{1999}

\begin{document}

\maketitle

\label{firstpage}

\begin{abstract}  

We report the first IR proper motion measurements of the Herbig-Haro objects in
the Orion Molecular Cloud--One using a four-year time baseline. The [Fe\,II]
emitting bullets are moving of order 0.08 arcsec per year, or at about 170 \kms.
The direction of motion is similar to that inferred from their morphology. The
proper motions of \h2\ emitting  wakes behind the [Fe\,II] bullets, and of
newly found \h2 bullets, are also measured. \h2\ bullets have smaller proper
motion than [Fe\,II] bullets, while \h2\ wakes with leading [Fe\,II] bullets
appear to move at similar speeds to their associated bullets. A few instances of
variability in the emission can be attributed to dense, stationary clumps in the
ambient cloud being overrun, setting up a reverse--oriented bullet. Differential
motion between [Fe\,II] bullets and their trailing \h2 wakes is not observed,
suggesting that these are not separating, and also that they have reached a
steady--state configuration over at least 100 years. The most distant bullets
have, on average, larger proper motions, but are not consistent with free
expansion. Nevertheless an impulsive, or short--lived ($\ll$ 1,000 years)
duration for their origin seems likely.

\end{abstract}


\section{Introduction}

The Orion Nebula (M42, NGC 1976) is the nearest massive star forming region, at
a  distance of 450\,pc  \cite{Genzel89}, thus becoming the archeytpe for
studying related phenomena such as outflows, shocks and photodissociation.  A
score of planetary--mass `bullets' were discovered in the core of Orion 
Molecular Cloud--One (OMC--1) by \citeANP{AB} (\citeyearNP{AB}, AB hereafter). 
The bullets are prominent in near--IR [Fe\,II] $a^4 F_{5/2}$ -- $a^4 D_{7/2}$
1.64 $\mu$m line emission, with trailing wakes bright in H$_2$ 1-0 S(1) 2.12
$\mu$m line emission, which are sensitive to intermediate ($\geq$ 50 \kms) and
slow ($\leq$ 50 \kms) speed shocks, respectively. The bullets were identified
with fast moving optical Herbig--Haro (HH) objects observed by \citeN{Axon84}
in the [O\,I] 6300\AA\ line. As indicated by their line widths, they are
travelling out of the cloud at speeds of up to 400 \kms. Projecting back, the
wakes point to an origin within 5 arcsec of the IRc2--complex. Their dynamical
age is no more than 1,000 years. A second finger system was found by
\shortciteN{Stolovy98} from HST/NICMOS observations, within $\approx$ 15 arcsec
of the IRc2--complex. This inner finger system emits in H$_2$ only, with no
[Fe\,II] emission at their tips.  These bullets are presumably interacting at
lower speeds with the medium, leading to a lower--excitation spectrum.

The formation mechanism for these bullet systems is not yet clearly 
understood. Two competing hypothesis are 1) one or multiple explosive events
fling off ejecta from the IRc2--complex (AB) and 2) {\it in situ}
Rayleigh--Tayler instabilities in time-variable winds, or in a slow wind being
overtaken by faster moving wind \shortcite{Stone95}.  A proper motion study can
verify the bow shock interpretation of the bullet-wake systems. If motions are
not detected, perhaps because the emission arises in a standing wave in the
flow of a wind from a central source, the ejection model (AB) would be
invalid.   Together with line profile measurements, proper motions may
constrain the location of the central source by determining the  three
dimensional structure and velocities. In addition,  relative proper motions of
the H$_2$ and [Fe\,II] gas in a wake and bullet provide constraints on how
ambient material is  entrained by the leading bullet.

\begin{figure*} 
\vspace{-0.7cm}
\begin{tabular}{c}
 \hspace{-2.4cm}     
 \psfig{figure=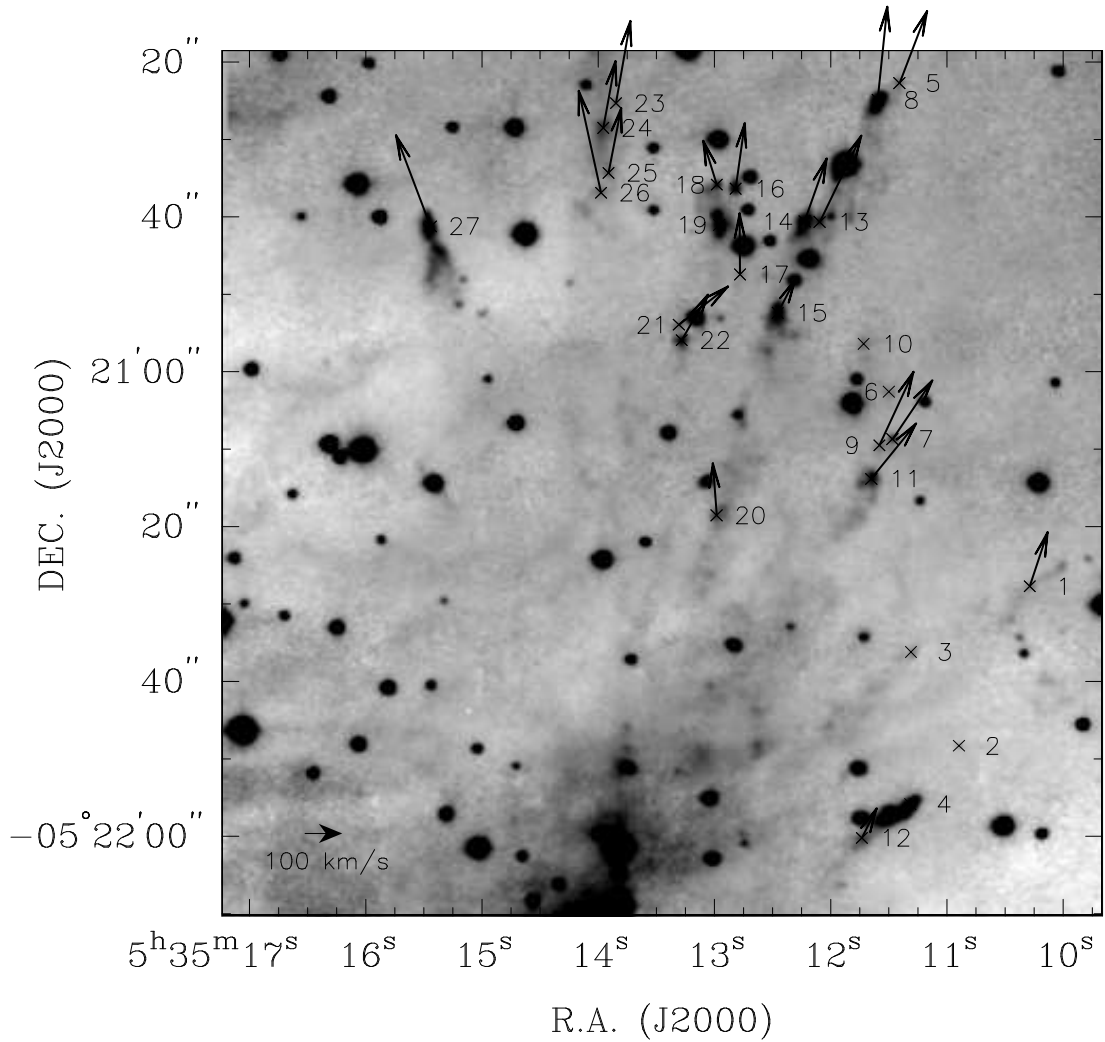,width=18cm,angle=0}\\
 \vspace{-11.2cm}
\end{tabular}

\caption{The [Fe\,II] 1.64 $\mu$m image of OMC--1 taken in December 1996, with
proper motion vectors from the centroid method overlaid. For HH\,117-200
(No.\,12) and HH\,130-119 (No.20) the proper motion vectors from eye--estimation
are used (see $\S$3.1). The length and orientation of an arrow represents the
transverse speed of a bullet and its direction of motion.  A scale bar in the
lower left corner corresponds to a speed of 100 \kms\ at a distance of 450 pc
and its length to the distance a bullet would cover in 100 years. The
IRc2--complex, where the driving source is presumed to be located, is at $5^h
35^m 14\fs 4$, $-5^{\circ} 22' 30''$ (J2000), below the image. Every bullet
whose proper motion has been measured is numbered (first column of Table 1).
Those whose motion has S/N $<$ 3 are marked with a cross only. }

\label{fig:feii_pm}
\end{figure*}

\begin{figure*}  
 \vspace{-0.7cm}  
 \begin{tabular}{c} 
 \hspace{-2.4cm}  
\psfig{figure=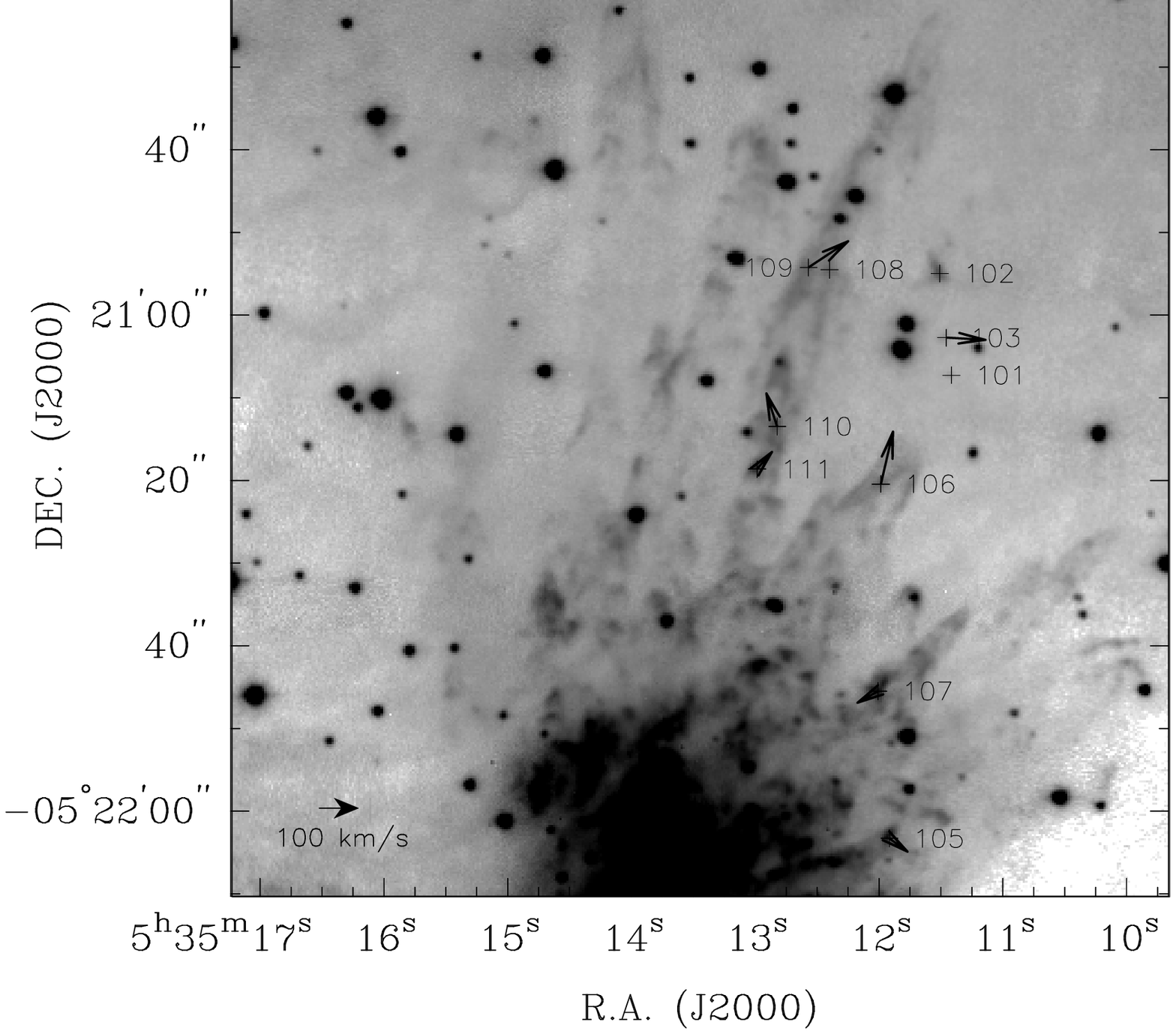,width=18cm,angle=0}\\
 \vspace{-11.2cm} \end{tabular}

\caption{The \h2 1--0 S(1) 2.12 $\mu$m image of OMC--1 taken in December 1996,
with proper motion vectors overlaid as Figure \ref{fig:feii_pm}. A newly found
\h2 bullet HH\,117-212 (No.\,104) is out of the frame and not shown. For its
location, see Figure \ref{fig:h2bullet}. Every \h2 knot whose proper motions
have been measured is numbered (first column of Table 1). Those with S/N $<$ 3
are marked with a plus sign (+) only.}

\label{fig:h2_pm}
\end{figure*}

Previously \citeN{Jones85} measured the proper motion of three optical bullets
(HH\,201, 205 and 210, following the nomenclature of \citeNP{Reipurth99}) by
comparing 20 plates, all taken on infrared emulsion (1--N or 103a--U) through an
RG-8($\equiv$ RG--695) filter with the Lick Observatory 120-inch telescope
between 1960 and 1983. They showed large tangential velocities, and the authors
speculated that the sources IRc2 and IRc9 must be responsible for the
excitation, residing at most $\sim$0.2 pc behind an obscuring molecular cloud.
\citeN{Hu96} used narrow band [N\,II] 6583\AA\ images taken by HST/WFPC with a
4 year baseline to also determine their proper motions, and  long-slit, high
resolution echelle spectra of [O\,I] in measuring the three dimensional
kinematics. He estimated shock velocities and orientation angles for HH\,201,
205, 206 and 210 adopting the bow shock model and analysis by
\shortciteN{Hartigan87}. He suggested that a hydrodynamic instability in a shell
swept up by a poorly collimated stellar wind from young stars is responsible
for the formation of these objects. 

Through high resolution spectroscopic observations of the [Fe\,II] 1.64 $\mu$m
line,  \shortciteN{Tedds99a} obtained line profiles for the bullets HH\,126-053
(=HH\,207; here HH\,125-053, see text) and HH\,120-114 (here HH\,117-114). From the shape
of the integrated [Fe\,II] line profiles, they derived the bullet velocity and
the orientation angle based on the analysis of \shortciteN{Hartigan87} that
showed the full-width zero intensity (FWZI) of the integrated
profile over a single bow shock must be equal to the speed of the bullet
itself. The results from these studies are give in Table 1 for comparision. We
have used images of the [Fe\,II] and \h2 lines, taken during periods of
exceptional seeing and with a 4--year baseline, to investigate proper motions
of the multitude of bullets which are observable at near--IR wavelengths.

Data acquisition and analysis is presented in Section 2, with the results from
this and previous studies in Section 3.  Discussion and plans for future
work are followed in Section 4 and 5, respectively.

\section{Observations \& Data Analysis}

A region of $\sim$ 2 $\times$ 2 arcmin in OMC--1 was mapped in 1992 September 13 and
1996 November 24, giving a 4.2 year time baseline, with the IRIS near--IR camera
\shortcite{Allen93}  on the 3.9-m Anglo-Australian  Telescope. Narrow-band line
images in [Fe\,II] 1.64 $\mu$m and H$_2$  2.12 $\mu$m emission were taken with 30
second integration time.  A pixel scale of 0.27 arcsec was used for both data sets
with  sub-arcsecond seeing (0.7\arcsec\ in 1992 and 0.9\arcsec\ in 1996) for both
epochs. The data reduction process involved the usual bad pixel removal,
flatfielding and sky subtraction.  A slight field rotation (12.2$'$) between the two
epochs was corrected for the second epoch data with a transformation solution
obtained by using common stars in the field.  The error involved in image
registration is of order a tenth of a pixel. We did not attempt to correct for
different seeing between the epochs.

Since the bullets and trailing wakes are diffuse objects, their proper motion
depends on whether one is measuring their bulk motion, and/or viewing changes
in morphology as well.  We adopted three  methods which are sensitive to
different aspects of motion; 1) image centroiding, 2) cross--correlation and 3)
eye--estimation. Centroiding is sensitive to motion of the emission centroid of
an object, and cross--correlation to morphological changes. A small area of sky
around an object is selected in two epochs and cross--correlated, producing a
resultant image. The amount of shift of the peak intensity from the centre in
this cross--correlated image was interpreted as proper motion. In reality,
however, it reflects both bulk motion and changes in shape of the object. If
the results from the first two methods do not agree, we attributed  this to
morphological changes, which was often confirmed by the third method -- eye 
examination. Contours of the object in two epochs were compared by eye to give
the shift in the position of intensity peaks only. 

In naming a bullet, we followed the nomenclature proposed by 
\shortciteN{Odell94}  to designate compact sources and stars in M42. It assigns
to an object a catalog number indicating the position of the source.  The first
three digits indicate the position in right ascention (E2000), and the second
three  the position in declination.  The common values for the inner region of
M42 (5$^h$35$^m$ and $-5^\circ 20'$) are not included. Hence, a bullet located
at $5^h 35^m 11\fs 51$, $-5^\circ 21' 47''$ would become 115-147.   The
coordinates of PC043, $5^h 35^m 13\fs 39$, $-5^\circ 21' 08.0''$,  from a HST
WF/PC survey \shortcite{Prosser94} was used as the reference position to obtain
the coordinates of bullets.

\begin{figure*} 
\begin{tabular}{c}
\psfig{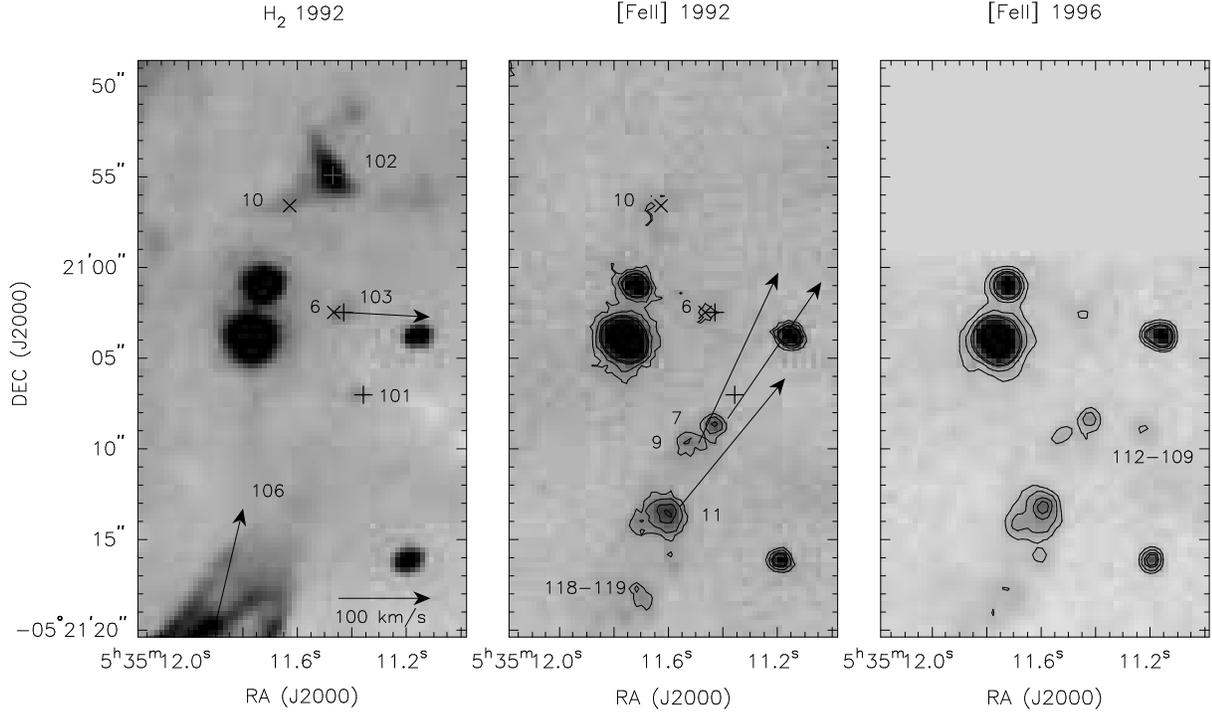}\\
\end{tabular}  

\caption{Enlargened images around HH\,117-114 of \h2 in 1992 ({\it left}),
[Fe\,II] in 1992 ({\it centre}) and [Fe\,II] in 1996 ({\it right}).
The compact [Fe\,II] emission knot, HH\,117-114 (No.\,11), lies at the
head of wakes in H$_2$. Two more [Fe\,II] bullets, HH\,115-109 (No.\,7) and
HH\,116-110 (No.\,9) are in front of it. A stationary clump 114-107
(No.\,101) in H$_2$ emission lying ahead of them. Presumably it is being rammed
by the HH\,115-109/HH\,117-114 (Nos.\,7/11) bullet system. An [Fe\,II] clump shows
up as HH\,112-109 only in 1996, and was not seen in 1992 (middle and right
panels). HH\,117-057 (No.\,10) is stationary,
accompanyed by a bright triangle-shaped stationary \h2 clump, 115-055
(No.\,102) next to it.  The velocity vector and the scale bar, on the left
 panel, is the same as in Figure \ref{fig:feii_pm}. \h2 knots are marked
by + and  [Fe\,II] knots by $\times$.}

\label{fig:117}
\end{figure*}

\section{Results}

The results of the proper motion measurements of the Orion bullets and wakes
are listed in  Table\,1 (see also Figure \ref{fig:feii_pm} for [Fe\,II] and
Figure \ref{fig:h2_pm} for \h2), as well as the results from  previous
studies.  The average proper motion of [Fe\,II] bullets and their trailing \h2\
wakes is about 8 arcsec per century for those whose proper motion has been
measured, corresponding to a transverse speed of 170 \kms\ at the  distance of
450\,pc  to OMC--1. For bullets whose motion has S/N $\geq$ 3, the average
proper motion is 9.4 arcsec per century with a corresponding transverse speed
of 200 \kms. The direction is similar to that inferred from their morphology. 
In general, the three methods have given similar results (aside from
HH\,109-148, HH\,113-156 and HH\,130-041) as there are few clear changes in the
bullet morphology over the 4 year baseline. However some bullets and wakes have
change in shape and/or intensity, as described below. We first discuss  the
proper motions seen in the [Fe\,II] images, and then in the \h2 images.

\subsection{Noticeable Features in [Fe\,II] Line Emission} 

Features for individual [Fe\,II] bullets are described below in the ascending
order of their R.A. coordinate. Their designation from the previous studies are
also given in parenthesis when available, and the number following is from
Column 1 in Table 1.\\

\noindent {\bf  HH\,113-156 (=HH\,201) \& HH\,117-200} [Nos.\,4 \& 12]
 
This shows a double feature and/or unresolved substructure within it, $\sim 
1.5$\,arcsec, or $\sim 600$\,AU  in size. \citeN{O'Dell96} identified this as
four seperate objects of 113-153, 114-155, 155-155 and 116-156. It has no
counterpart or wake in \h2, attributed to its motion close to our line of
sight. Different methods give different proper motion values, mainly due to its
double feature and unsymmetric shape. The tangential velocity $v_{tan}$
calculated from centroiding, $38 \pm 21$ \kms, disagrees with those from
previous studies. This probably results from partially resolving the feature
into two components.  $v_{tan}$ obtained from cross--correlation (177 \kms) and
from eye--estimation (100 \kms) both agree reasonably. They give an orientation
angle (O.A.) of 60\degr\ for the former, and 20\degr\ for the latter method when
combined with \citeN{Hu96}'s estimation of the shock velocity 310\,\kms. O.A.
is the angle between the direction of motion of an object in space and our
line--of--sight. 

HH\,117-200 (No.\,12) shows weak [Fe\,II] emission, located just southeast of
HH\,113-156 (No.\,4), however their association is not clear. The proper motion
determined for HH\,117-200 (No.\,12) vary for different methods; centroiding
gives a motion to the northeast with P.A.=30\degr\ while the other two methods
to the northwest with P.A.=330\degr.  Its $v_{tan}$ also varies from 90 \kms\
(eye--estimation) to 213 \kms\ (cross--correlation). Its overall intensity
distribution remains the same for both epochs, and a small shift in emission
peak is attributed as the  cause of this discrepancy.\\

\noindent {\bf  HH\,115-109/HH\,117-114 (=HH\,120-114)} 

\hfill[Nos.\,6, 7, 9, 10 \& 11]

This bullet system was designated as HH\,120-114 by \shortciteN{Tedds99a}
(without exact astrometry) using AB's 1992 image, but we seperate them into two
(possibly three or more, see below) bullets with clear evidence of a double
feature in the \h2 wakes behind them (Figure \ref{fig:117}). The bullets are
moving at a similar speed of $\sim$200 \kms\ and position angle of 320\degr.
HH\,116-110 (No.\,9), a secondary emission peak in the [Fe\,II] wake of HH\,115-109
(No.\,7), moves slightly faster than HH\,115-109 (No.\,7). It is not clear
whether they are  seperate bullets. \shortciteN{Tedds99a} derived the bullet
velocity and O.A.\ for  this system from the shape of the integrated [Fe\,II]
profiles. They derived a shock velocity $v_s = 120 \pm 10$ \kms\ and O.A.\ =
$60 \pm 15$\degr\, giving a corresponding $v_{tan}$ of $100 \pm 25$ \kms. The
difference in $v_{tan}$ from that derived here indicates that using the shape
of line profiles to infer orientation angles can be unreliable. In this
particular case several profiles are overlapping.

A knot at 115-103 (No.\,6/103) shows weak emission in both [Fe\,II] and \h2
lines with a slight offset. Both species show similar proper motions of 100
\kms. Note the proper motion in [Fe\,II] from eye--estimation is greater than
that from centroiding, due to a small shift in emission peak to
west. HH\,117-057 (No.\,10) is found to be stationary next to a bright \h2 clump
115-055 (No.\,102).  From the relative locations of [Fe\,II] and \h2, we
suggest that this is a `reverse' bullet, i.e.\ a dense, stationary condensation
being rammed by outflowing material, producing a reverse-oriented bow shock
(see $\S$4).

The proper motion of a knot at 118-119 was not measurable because it lies in
the middle of diffuse [Fe\,II] emission in the tail of the
HH\,115-109/HH\,117-114 (Nos.\,7/11) system. In addition the weak emission peak
in 1992 had become more diffuse by 1996. Interestingly it lies along the axis
of one of the following \h2 wakes, suggesting this is more directly associated
with them (see $\S$3.2). Note an [Fe\,II] clump which showed up at 112-109 in
1996 (Figure \ref{fig:117}).  It was not seen in 1992 indicating 
intensity variability. \\

\begin{figure}
\vspace{-0.6cm} 
\begin{tabular}{c}
\psfig{figure=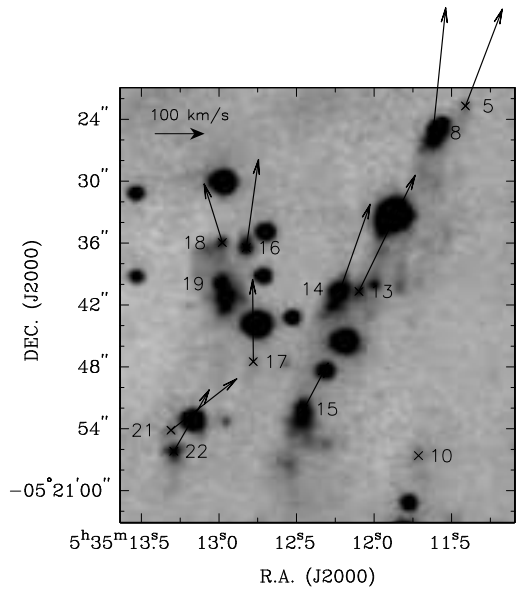,angle=0,width=1\textwidth}
\end{tabular}
\vspace{-16.8cm}

\caption{A blow-up of the most crowded field accomodating HH\,116-025,
HH\,122-041 and HH\,125-053 (Nos.\,8, 14 \& 15). The velocity vector and the
scale bar, at the upper left corner, are the same as in Figure
\ref{fig:feii_pm}.}  

\label{fig:HH567}
\end{figure}

\noindent {\bf HH\,116-025, HH\,122-041 \& HH\,125-053
(=HH\,205, HH\,206 \& HH\,207)} [Nos.\,5, 8, 13, 14 \& 15]

This is the most crowded field with several bullets and wakes overlapping
(Figure \ref{fig:HH567}). Thus this field is suitable for determining whether
any differential motion between bullets and/or between species occurs, i.e.\
whether or not the bullets move with different speeds depending on their
distance from the origin, and/or if their \h2 wakes are stretching out behind
them. However, no differential motion was detected. This may be attributable to
the short time baseline. If such motion was, for example, 30 \kms, a 10-year
baseline would be required.

HH\,116-025 (=HH\,205, No.\,8) is one of the fastest moving bullets ($\sim$
250\,\kms) at the tip of the most crowded finger. \citeN{Hu96} reported a
morphological change in [N\,II], but no change is apparent in our [Fe\,II]
image. His estimation of $v_s$ = 310\,\kms\ with O.A.\ = 90\degr\ (moving on
the plane--of--sky), makes the expected tangential velocity the same as the
shock velocity.  A smaller bullet, HH\,114-023 (No.\,5), located further
northwest of it, shows similar proper motion.   HH\,122-041 (No.\,14) is
irregular in shape having the same direction of motion with HH\,116-025 and 
HH\,125-053 (Nos. 8 \& 15). For HH\,125-053 (=HH\,207, HH\,126-053; No.\,15),
$\sim 100$\,\kms\ proper motion is measured while the estimation by
\shortciteN{Tedds99a} gives $v_{tan} = 140 \pm 25$\,\kms, with $v_s = 150 \pm
10$\,\kms\  and O.A.\ $=70 \pm 15$\degr. Emission at the bullet head is
observed to get stronger without any morphological change. \\

\noindent {\bf HH\,130-119} [No.\,20]

HH\,130-119 is located along the line of a series of prominent bullets;
HH\,116-025, 122-041 and 125-053 (Nos.\,8, 14 and 15, respectively). It is quite
diffuse in shape with no noticeble emission peak in the 1992 image. In 1996,
however, an emission peak was prominent in the lower-left corner of the bullet,
considerably affecting the proper motion measurement using image centroid and
cross--correlation. By blinking between two epoch data, we can see its motion
toward northwest (P.A.=5\degr), and this value is used in Figure 1. This
demonstrates  the limitation of centroiding  when a bullet changes its
intensity distribution over time. Apparently it has the largest proper motion, 
but this is  a reflection of the change in distribution within its diffuse,
irregular morphology, rather than of actual bulk motion.\\

\begin{figure} 		
\psfig{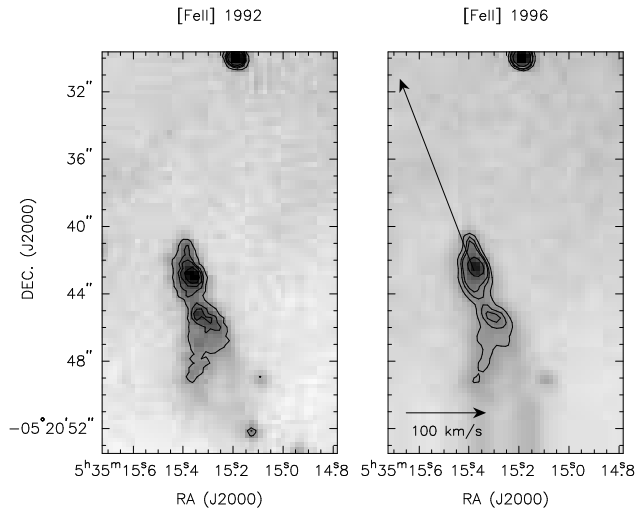}

\caption{HH\,154-041 (=HH\,210, No.\,27) is shown in [Fe\,II] line emission with
the proper motion vector overlaid. Two or more bow features overlap here. While
the overall shape is preserved, the lower emission peak has slightly shifted
to the right, in comparison to the motion of the bullet head. The velocity
vector and the scale bar, on right panel, are the same as in Figure 1.}  

\label{fig:210}
\end{figure}		

\noindent {\bf HH\,154-041 (=HH\,210)} [No.\,27]

\citeN{Jones85} reported a `nucleus' in its head. We also see this in [Fe\,II]
and in J band (1.25\mic), as much fainter emission with similar morphology,
but not in \h2 emission. This could mean that its motion is close to our line
of sight, or that it has escaped from the molecular cloud, hence no \h2. While
zero proper motion was expected from \citeN{Hu96}'s estimation of $v_s = 400$
\kms\ and O.A.\ = 0\degr\ (moving directly toward us), our measurement gives
$v_{tan}$ = 260 \kms. In fact, the morphology suggests there are at least two
bow features overlapping -- while the overall shape is preserved the second
emission peak shows a small shift (0.6$''$) 
to southwest, different from the northeast motion of the tip of the bullet.\\

\subsection{Discovery of New \h2\ Bullets} 

\begin{figure} 
\psfig{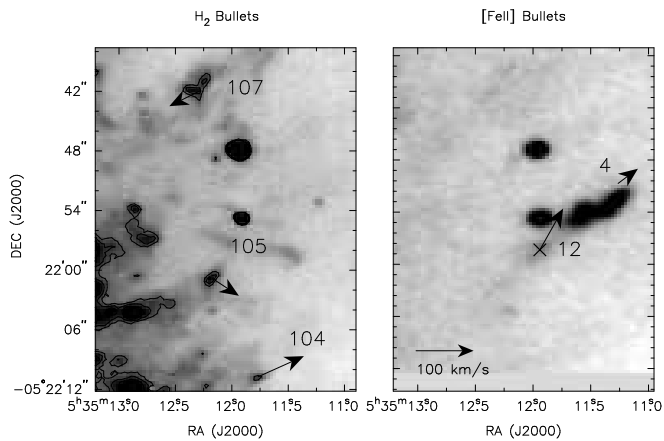}

\caption{({\it left}) Newly discovered \h2 bullets  are shown with proper
motion vectors overlaid as Figure 1.  They are found between HH\,113-156
(=HH\,201, No.\,4; {\it right}) and the presumed origin in the IRc2--complex.
The corresponding [Fe\,II] line emission is shown right. The proper motion
vector for HH\,113-156 (No.\,4) is offset for clarity, and the proper motion
vector from the eye--estimation is used for HH\,117-200 (No.\,12; see text).
The velocity vector and the scale bar, on right panel, are the same as in
Figure 1.}  

\label{fig:h2bullet}
\end{figure}

We report the discovery of new \h2 bullets; HH\,117-212, HH\,119-203 and
HH\,120-145  (Nos.\,104, 105 and 107, respectively; Figure \ref{fig:h2bullet}).
They emit in \h2\ only, located between HH\,113-156 (=HH\,201, No.\,4) and the
presumed origin in the IRc2--complex. Their morphology resembles that of
[Fe\,II] bullets but with smaller proper motions of order $\sim$100 \kms. We
presume their slower speed results in a lower-excitation spectrum without
[Fe\,II]. These are thus similar to the inner system of \h2 bullets discovered
within 15 arcsec of IRc2--complex by \shortciteN{Stolovy98}.

\subsection{Noticeable Features in \h2 Line Emission} 
  
The proper motion measurements for both \h2\ emitting wakes and \h2 bullets
were difficult to determine because of their diffuse morphology. H$_2$ wakes
which do have leading [Fe\,II] bullets appear to have a similar proper motion
vector as their leading bullet. However, in general, the proper motions of
H$_2$ bullets are slower than those of [Fe\,II] bullets. \\

\noindent {\bf HH\,114-107, 115-055 \& 115-103} [Nos.\,101, 102 \& 103]

HH\,114-107 (No.\,101) at the tip of HH\,115-109/117-114 (Nos.\,7/11) has no
associated [Fe\,II] emission (Figure \ref{fig:117}) and is found to be
stationary. This may be a stationary cloudlet being rammed by the shock wave
running ahead of the HH object, as is the case with the `tadpoles' in Helix
Nebula (eg.\ \shortciteNP{Burkert98}). In such a case, the nearside of the
clump to the IRc2--complex would show the highest excitation with its trail
pointing away from the complex. Unfortunately the size of the knot is not big
enough for such structure to be resolved. 

The same explanation can be applied to 115-055 (No.\,102) which is a bright
triangle-shaped clump with a weak [Fe\,II] emission knot HH\,117-057 (No.\,10)
interior to it. When their emission peaks in [Fe\,II] and \h2 are connected,
the line has P.A.\ of $\sim 310$\degr, pointing about 30 arcsec  east of the
IRc2--complex. 

For 115-103 (No.\,103), the emission peak has shifted to west by half
an arcsecond giving much larger proper motion from eye--examination than from
centroiding, as has its counterpart in [Fe\,II] emission HH\,115-103 (No.\,6). \\


\noindent {\bf 120-120} [No.\,106]

\begin{figure} 
\psfig{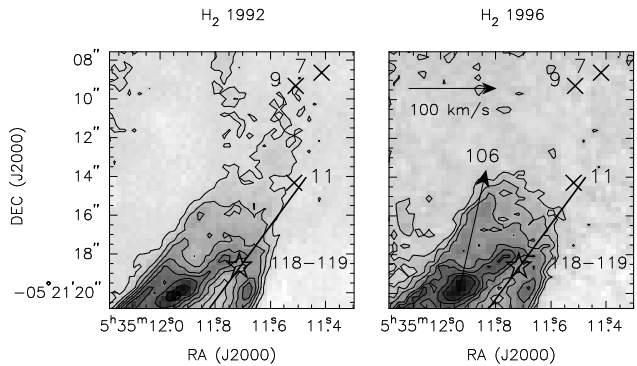}

\caption{The open head of wakes accomodating \h2 clump 120-120 (No.\,106), lead
by the [Fe\,II] bullet system HH\,115-109/HH\,117-114 (Nos.\,7/11). A straight
line is drawn along the presumed axis of the smaller bow feature with P.A.\ of
320\degr. A star sign marks the position of a [Fe\,II] knot 118-119, and
$\times$ the  [Fe\,II] bullets HH\,115-109 (No.\,7), HH\,116-110 (No.\,9) and
HH\,117-114 (No.\,11). The velocity vector and the scale bar in the right panel
are the same as in Figure 1. See Figure \ref{fig:117} for the proper motion
vectors for the [Fe\,II] bullets shown here.}  

\label{fig:117h2}
\end{figure}

Two bow features are discernable in \h2 wakes associated with an emission peak
at 120-120 (No.\,106) (Figure \ref{fig:117h2}). They are trailing the [Fe\,II]
bullets HH\,115-109 (No.\,7) with P.A.=330\degr\ and HH\,117-114 (No.\,11) with
P.A.=320\degr.  A line, drawn along the presumed axis of the smaller bow
feature, passes through a secondary [Fe\,II] emission peak at 118-119 which
coincides with the tip of the bow. Extended further northwest, the line passes
through HH\,117-114 (No.\,11). These features may arise from the fragmentation
of the bullet into two pieces, rather than these being two seperate bullets
projected on top of one another. 

Figure \ref{fig:117h2} also shows a morphological change at the head of these
wakes. The cone-shaped tip of the \h2 wake in 1992 has disappeared leaving it
flat-headed by 1996. Inhomogeneity in the ambient molecular cloud on scales of
$\sim$ 100 AU, causing changes in the rate of
entrainment of material behind a bullet,  could produce such a change.



\section{Discussion}			

The most basic result from the data presented here is that the bullet/bow-shock
interpretation placed on the observations of the [Fe\,II] caps and H$_2$ wakes
in Orion by AB is essentially correct.  The proper motions
measured confirm that the bullets are travelling away from the core of the
cloud at supersonic speeds, and are thus projectiles that are being shocked. 
They cannot be stationary cloudlets, nor can they be standing waves in a flow. 

Their speeds, typically of order 200 \kms, and directions are consistent with
both the widths of their line profiles, and with the morphology of individual
bullets and fingers.  We are observing bullets and associated wakes interacting
with an ambient medium, spread over a wide opening angle, as they move outwards
from the centre of the molecular cloud. 

The inferred speeds are much in excess of the dissociation speeds for H$_2$ and
even [Fe\,II], and thus cannot be the actual shock speed of the line
emitting gas.  There is, however, no suggestion from other data (eg.\
\shortciteNP{Tedds99b}) that the bullets are
traversing a moving medium (thus reducing the shock velocity to the difference
between the bullet and medium velocities).  Oblique shocks, except at the very
head, must occur with a significant tangential component of velocity remaining
in the shocked material.  This is as expected for a bow shock around a bullet,
with only the normal component of the bullet's motion at each point along the
bow contributing to the shock speed. 

We see little evidence for significant morphological changes in the bullets
over the four-year baseline, nor for fading or brightening of individual knots
over the same time, aside from [Fe\,II] clumps HH\,112-109 and HH\,130-119
(No.\,20), and the head of the \h2 wake associated with 120-120 (No.\,106) (see
$\S$3).  This is thus different from the observation of some of the bullets in
HH46/47 \shortcite{Micono98} and HH\,111/121 \shortcite{Coppin98}, where
brightness variations on timescales of 1 year (the cooling time for H$_2$) are
seen. Such variations likely arise from inhomogeneities in the ambient medium
over spatial scales of at most some tens of AU, leading to changes in the
amount of material being swept up over this time.  For the bullets in Orion the
medium must thus be reasonably homogeneous over $\sim 100$ AU scales, otherwise
changes in shock strength, and thus line intensities, would be observed.
However,  the change in \h2 wake intensity near 120-120 (No.\,106) and
appearance of [Fe\,II] clump HH\,112-109 (see Figure \ref{fig:117}) suggests
this region is not homogeneous over this scale. Here, a bullet may have
fractured into several pieces, each forming its own set of \h2 wakes, overlapping 
behind. The similar transverse speeds of [Fe\,II] bullets HH\,115-109,
HH\,116-110 and HH\,117-114 (Nos.\,7, 9 \& 11) support this suggestion.


No differential motion, however, between [Fe\,II] bullets and their H$_2$ wakes
is observed.  While limits on such motion remain weak over a four year
timescale, this suggests that the bullets are not separating from their wakes.
Together with the lack of any morphological changes, the wakes also cannot be
stretching.  In other words, each bullet and wake appears to be moving
together.  This does not, however, imply that the gas within the bullet/wake is
stationary with respect to the bullet velocity.  Material swept-up by the
bullet will be swept down the wake at the tangential component of the shock
velocity, and will thus traverse the length of the bow.  It will continue to be
observed until it reaches the Mach point, where the normal component of the
bullet velocity drops below the minimum shock speed to excite the gas ($\sim 5$
\kms\ for molecular gas).  The time taken for this will be of order 100 years. 
Since the wakes are not stretching either, the bullets must also have been in
motion for at least this long, in order to reach a steady-state configuration. 

The bullets observed in H$_2$ only, on the inner edge of the finger system
HH\,113-156 (No.\,4), are moving more slowly ($\sim 100$ \kms) than those
emitting in [Fe\,II] ($\sim 200$ \kms), though their speeds still exceed that
for dissociation of H$_2$.  However, these slower speeds are consistent with
lower excitation for the inner bullets, as would be inferred by the absence of
[Fe\,II] in them.  This gives support to the inner H$_2$ bullet system
discovered by Stolovy et al.\ (1998) within 15 arcsec of IRc2--complex as
being slower moving analogues of the [Fe\,II] bullets. 

Nevertheless not all line-emitting knots are moving outwards at these speeds. 
For instance, HH\,117-057 (No.\,10) and 115-055 (No.\,102) appears to be
stationary with respect to the ambient cloud. HH\,114-107 (No.\,101) is also
stationary
with HH\,115-109/HH\,117-114 (Nos.\,7/11) moving towards it. We hypothesise
that it is a dense clump in the ambient medium  just now being rammed by 
HH\,115-109/HH\,117-114 (Nos.\,7/11). Time variability is evident in 
[Fe\,II] clump, HH\,112-109 in Figure \ref{fig:117}, which showed up in
1996 but was not seen in 1992. In all these cases, we expect the clumps to 
brighten over the next few years as they are further overrun. 

The proper motion vectors of the bullets clearly extend back towards an origin
somewhere in the vicinity of IRc2--complex.  However their extrapolation
remains no more accurate than that undertaken by Allen \& Burton (1993), who
extended the wakes back towards an origin within 5 arcsec of the
IRc2--complex.  Our present data still cannot shed any more light on where this
might be, which awaits a longer time baseline of measurements. 

\begin{figure} 
\centering
\begin{tabular}{c}
\psfig{figure=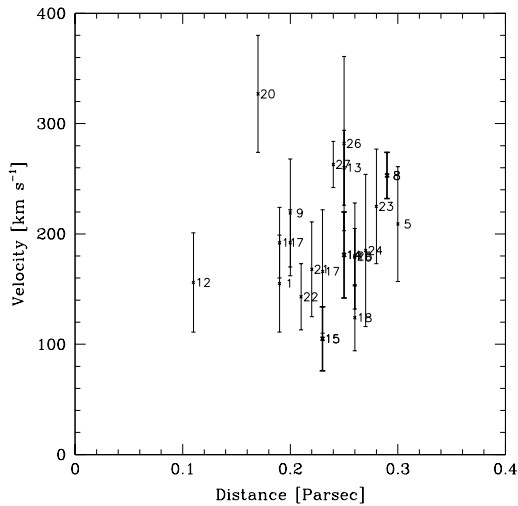,angle=0,width=7cm}\\
\end{tabular}

\caption{The bullet speeds whose S/N $\geq$ 3 are plotted against their
projected distance from the IRc2--complex, again numbered with those in the
first column of Table 1.  Thick solid lines are for HH\,116-025 (No.\,8),
122-041 (No.\,14) and 125-053 (No.\,15) from right to left (see $\S$4).}   

\label{fig:dv}
\end{figure}

The projected distance of bullets from IRc2--complex is plotted against its
proper motion speed in Figure \ref{fig:dv}. Excluding HH\,130-119 (No.\,20)
which was problematic in its measurement, the plot suggests that more distant
bullets, in general, are moving faster. This is consistent with an explosive,
or at least an impulsive event, as the origin for the bullets. Certainly it
must have occurred over a timescale much less than that of the expansion,
i.e.\ $\ll$ 1,000 years. This is most clearly demonstrated for HH\,125-053,
122-041 and 116-025 (No.\,8, 14 and 15, respectively) along the main finger,
which show a linear increase in speed with distance. However the motions are
not consistent with free expansion from the IRc2--complex; extrapolation to
zero speed would suggest an origin 0.05 -- 0.1 pc from the IRc2--complex. This
might correspond to where the bullets have formed, through instabilities in a
time variable wind.

\section{Future Work}

Our data represent just two epochs of observation, and is only the start of an
investigation of the motion of the Orion bullets.  With further epochs of data
we will be able to address further questions regarding their origin and
excitation, including: 

(a) whether the bullets are decelerating (changing in velocity)?

(b) are there non-radial motions, perhaps imparted by collisions with dense
objects? 

(c) are there more differential motions between the bullets and their wakes? 

(d) could we be seeing a pattern speed of a bullet, rather than its bulk
motion?  This may be evident if wiggles are observed on a proper motion
vector. 

(e) are there more morphological or intensity changes in time, such as the
stretching of a wake or the brightening or fading of a knot? These might arise
from inhomogeneities in the ambient medium, thus indicating what scale these
occur on. 

Further proper motion vectors will help tie down the origin of the driving
source, by reducing the errors on their extrapolation back to an origin. 
Combining the data with line profile measurements would give the 3D velocity of
a bullet, and allow a 3D picture of their distribution in space to be built
up.  The line profiles could then be compared to models such as those of
\shortciteN{Hartigan87} which predict profile shapes for a range of orientation
angles.  In addition, determining an orientation directly will help constrain
shock models for the line production, such as the form of the cooling law and
shape of the bow, which significantly effect the resultant profile.  The
[Fe\,II] data can also be compared to optical data obtained with HST, which
should yield more accurate velocities on shorter time base lines due to the
superior spatial resolution that can be obtained.

\begin{table*}
\centering
\begin{minipage}{178mm}
 \caption{Proper Motions of the Orion Bullets and Wakes in [Fe\,II] and \h2 emission.}
 \begin{tabular}{rrrrrcrcp{18mm}l}
 \hline
 &	&	& 	& This study   &	&	&
 \multicolumn{2}{c}{Previous studies} &\\
 &Bullet& \multicolumn{1}{c}{$\mu^1$}  & $v_{tan}^2$ 
 	& \multicolumn{1}{c}{P.A.$^3$} & \multicolumn{1}{c}{$\mu^4$} 
 	& \multicolumn{1}{c}{$\mu^5$}  & $v_{tan}^2$, P.A.$^3$   & $v_{s}$, O.A.$^6$&\\
No. &Designation & ("/cent.)  & (\kms) & \multicolumn{1}{c}{(deg)} & ("/cent.) & ("/cent.) 
        & (\kms, deg)  &(\kms, deg) &Remark\\
 \hline\hline
\multicolumn{10}{l}{\bf [Fe\,II] knots} \\
 1& HH\,103-128  & 7.2 $\pm$ 2.1 & 155 $\pm$ 44 & 342 $\pm$ 16 & 6.4 &  6.4 $\pm$ 0.4 & &&\\
 2& HH\,109-148  & 1.6 $\pm$ 2.4 &  34 $\pm$ 51 & 293 $\pm$ 56 & 1.9 &  8.0 $\pm$ 0.4 & &&\\
 3& HH\,113-136  & 5.1 $\pm$ 2.0 & 110 $\pm$ 43 & 132 $\pm$ 21 & 3.1 &  7.2 $\pm$ 0.4 & &&\\
 4& HH\,113-156  & 1.8 $\pm$ 1.0 &  38 $\pm$ 21 & 297 $\pm$ 29 & 8.4 &  4.9 $\pm$ 0.4 & 170, 324 $^a$ & 310, 15 $^a$& HH\,201\\
  & 	       &	       &              &              &     &                & 167, 304  $^b$& & \\
 5& HH\,114-023  & 9.8 $\pm$ 2.4 & 209 $\pm$ 52 & 339 $\pm$ 14 &10.2 &  8.5 $\pm$ 0.4 & &&\\
 6& HH\,115-103  & 4.6 $\pm$ 3.2 &  99 $\pm$ 69 & 257 $\pm$ 35 & 2.1 &  7.7 $\pm$ 0.4 & &&\\
 7& HH\,115-109  & 9.0 $\pm$ 1.4 & 192 $\pm$ 30 & 326 $\pm$ 09 & 8.5 &  8.6 $\pm$ 0.4 &           && HH\,120-114 $^c$\\
 8& HH\,116-025  &11.8 $\pm$ 1.0 & 253 $\pm$ 21 & 354 $\pm$ 05 &10.1 &  9.2 $\pm$ 0.4 & 253, 343 $^a$ & 120, 90
 $^a$ &HH\,205\\
  &	       &	       &              &              &     &                & 236, 344 $^b$ & & \\
 9& HH\,116-110  &10.3 $\pm$ 2.3 & 219 $\pm$ 49 & 335 $\pm$ 13 & 8.8 & 12.0 $\pm$ 0.4 & & &\\
10& HH\,117-057  & 3.4 $\pm$ 3.2 &  72 $\pm$ 68 & 213 $\pm$ 43 & 2.3 &  4.5 $\pm$ 0.4 & & &\\
11& HH\,117-114  & 9.0 $\pm$ 1.5 & 192 $\pm$ 32 & 321 $\pm$ 09 & 6.3 & 10.0 $\pm$ 0.4 &           & 120, 60 $^c$ &
HH\,120-114 $^c$\\
$^*$12& HH\,117-200  & 7.3 $\pm$ 2.1 & 156 $\pm$ 45 &  31 $\pm$ 16 &10.0 &  4.2 $\pm$ 0.4 & &&\\
13& HH\,121-041  &12.2 $\pm$ 1.6 & 260 $\pm$ 34 & 334 $\pm$ 08 &12.0 &  9.8 $\pm$ 0.4 & &&\\
14& HH\,122-041  & 8.5 $\pm$ 1.8 & 181 $\pm$ 39 & 341 $\pm$ 12 & 7.9 & 12.4 $\pm$ 0.4 & 251, 333 $^a$ & &HH\,206 \\
15& HH\,125-053  & 4.9 $\pm$ 1.3 & 105 $\pm$ 29 & 332 $\pm$ 15 & 4.3 &  3.6 $\pm$ 0.4 &          & 150, 70 $^c$ &HH\,207\\
  & 	       &	       &              &              &     &                &          & & HH\,126-053 $^c$\\
16& HH\,128-037  & 8.4 $\pm$ 1.2 & 179 $\pm$ 26 & 352 $\pm$ 08 & 8.4 & 11.3 $\pm$ 0.4 & & &\\
17& HH\,128-048  & 7.7 $\pm$ 2.6 & 166 $\pm$ 56 &   0 $\pm$ 19 & 6.1 &  6.3 $\pm$ 0.4 & & &\\
18& HH\,130-036  & 5.8 $\pm$ 1.4 & 124 $\pm$ 30 &  17 $\pm$ 13 & 3.7 &  2.7 $\pm$ 0.4 & & &\\
19& HH\,130-041  & 1.6 $\pm$ 1.6 &  34 $\pm$ 34 & 254 $\pm$ 45 & 2.3 &  6.3 $\pm$ 0.4 & & &\\
$^*$20& HH\,130-119  &15.3 $\pm$ 2.5 & 327 $\pm$ 53 & 125 $\pm$ 09 & 5.6 & 6.6 $\pm$ 0.4 & &&  \\
21& HH\,133-054  & 7.9 $\pm$ 2.0 & 168 $\pm$ 43 & 308 $\pm$ 14 & 3.7 &  7.2 $\pm$ 0.4 & &&\\
22& HH\,133-056  & 6.7 $\pm$ 1.4 & 143 $\pm$ 30 & 330 $\pm$ 12 & 4.5 &  6.8 $\pm$ 0.4 & &&\\
23& HH\,138-026  &10.6 $\pm$ 2.5 & 225 $\pm$ 52 & 350 $\pm$ 13 & 6.9 &  7.6 $\pm$ 0.4 & &&\\
24& HH\,139-035  & 8.4 $\pm$ 2.2 & 180 $\pm$ 48 & 349 $\pm$ 15 & 6.9 & 12.6 $\pm$ 0.4 & &&\\
25& HH\,140-029  & 8.6 $\pm$ 3.2 & 185 $\pm$ 69 & 350 $\pm$ 20 & 8.4 &  7.7 $\pm$ 0.4 & &&\\
26& HH\,140-037  &13.2 $\pm$ 3.7 & 282 $\pm$ 79 &  12 $\pm$ 16 &11.9 &  7.0 $\pm$ 0.4 & &&\\
27& HH\,154-041  &12.3 $\pm$ 1.0 & 263 $\pm$ 21 &  21 $\pm$ 05 & 8.4 &  7.0 $\pm$ 0.4 & 215, 16 $^b$ & 400, 0 $^a$
&HH\,210\\
\multicolumn{10}{l}{\bf \h2 knots} \\
101& HH\,114-107 & 0.8 $\pm$ 1.6 &  17 $\pm$ 33 & 337 $\pm$ 62 & 1.7 & 0.0 $\pm$ 0.4  & & & stationary \\
102&   115-055 & 1.3 $\pm$ 1.1 &  27 $\pm$ 24 & 267 $\pm$ 41 & 3.3 & 2.4 $\pm$ 0.4  &  & & stationary?$^\#$ \\
103&   115-103 & 4.7 $\pm$ 1.8 & 100 $\pm$ 39 & 267 $\pm$ 21 & 5.4 &12.6 $\pm$ 0.4  &  & &  \\
104& HH\,117-212 & 4.2 $\pm$ 0.7 &  89 $\pm$ 15 & 333 $\pm$ 10 & 2.5 & 5.1 $\pm$ 0.4  &  & & New bullet\\
105& HH\,119-203 & 2.7 $\pm$ 0.7 &  57 $\pm$ 15 & 233 $\pm$ 15 & 3.3 & 3.3 $\pm$ 0.4  &  & & New  bullet\\
106&   120-120 & 6.3 $\pm$ 1.5 & 135 $\pm$ 32 & 347 $\pm$ 13 & 4.2 & 8.6 $\pm$ 0.4  &  & &  \\
107& HH\,120-145 & 2.8 $\pm$ 0.9 &  60 $\pm$ 19 & 150 $\pm$ 17 & 0.4 & 0.7 $\pm$ 0.4  &  & & New bullet\\
108&   124-055 & 3.1 $\pm$ 1.7 &  66 $\pm$ 36 & 299 $\pm$ 29 &     &11.6 $\pm$ 0.4  &  & &   \\
109&   126-054 & 5.6 $\pm$ 1.8 & 119 $\pm$ 39 & 304 $\pm$ 18 & 7.0 & 9.8 $\pm$ 0.4  &  & & \\
110&   128-113 & 4.1 $\pm$ 1.2 &  88 $\pm$ 25 &  18 $\pm$ 16 &     & 5.7 $\pm$ 0.4  &  & & \\
111&   130-119 & 2.7 $\pm$ 0.8 &  57 $\pm$ 17 & 319 $\pm$ 17 & 0.4 & 0.3 $\pm$ 0.4  &  & & \\
\hline
 \end{tabular}

\medskip 

The proper motion of bullets in \feii\ is followed by the proper motion of 
bullets and wakes in \h2 (from No.\,101 to 111). In naming \h2 knots, HH
indicating Herbig--Haro object is only given when it is an \h2 bullet, and not
when it is the wake behind an [Fe\,II] bullet.

\smallskip 
$^{1}$ From the shift in the intensity centroids of a bullet in two epochs. A
proper motion of 10 arcsec per century corresponds to a transverse speed of 215
\kms\ at the distance of 450\,pc to OMC--1.

\smallskip
$^{2}$ The tangential velocity; the velocity projected onto the plane--of--sky.

\smallskip 
$^{3}$   The position angle; the direction of motion in
the plane--of--sky, starting from N and increasing to E. 


\smallskip  

$^{4}$ From the cross--correlation of the distribution of emission from the
bullets at two epochs. It generally agrees with the results from the centroid
(see $\S$2). The diffuse shape of \h2 knots sometimes makes cross--correlation 
unreliable, when their value is not given.

\smallskip  
$^{5}$ From eye--estimation of the contour maps of bullets.
Eye--estimation is sensitive to changes in the intensity peak only. 

\smallskip  
$^{6}$ The shock velocity and orientation angle. 
The shock velocities are estimated from the line profile of [O\,I] 
\cite{Hu96} and [Fe\,II] \shortcite{Tedds99a}. The orientation angle is the
angle between our line-of-sight and the direction of motion of an object.
O.A.=90\degr\ when the object is moving on the plane--of--sky.

\smallskip    
$^*$ For these bullets, the proper motion vectors from the eye--estimation
(P.A.=333\degr\ for No.\,12 and  5\degr\ for No.\,20, respectively) are
used in Figure \ref{fig:feii_pm}  (see text).

\smallskip   
$^\#$ See $\S$3.2 for the discrepancy in the proper motion values from different
methods.

\smallskip
(a) \citeN{Hu96}, d=440 pc used; (b) \citeN{Jones85}, d=460 pc used;  
(c) \shortciteN{Tedds99a} 

\end{minipage}
\end{table*}

\section*{Acknowledgments}

We would like to thank Dr.\ J.\ A.\ Tedds for providing line profile
information for some of bullets studied here. This project is partly funded by
research grants from the `Small Grants' scheme of the Australian Research
Council (ARC).


\label{lastpage}

\end{document}